%% file: main.tex
\documentclass[onecolumn,journal,11pt]{IEEEtran}
\usepackage{amsmath,amsthm,amssymb}
\usepackage{enumerate}
\usepackage{color}
\usepackage{xcolor}
\usepackage{url}
\usepackage{balance}
\usepackage{graphicx} 
\usepackage{mathrsfs}
\usepackage{epsfig}
\usepackage{verbatim}
\usepackage{setspace}
\usepackage{cite}
\usepackage{multicol}
\usepackage{lipsum}
\usepackage{etoolbox}

\newtheorem{theorem}{Theorem}
\newtheorem{lemma}{Lemma}

\newtheorem{definition}{Definition}
\newtheorem{claim}{Claim}
\newtheorem{remark}{Remark}

\renewcommand{\vec}[1]{\mathbf{#1}}

\newcommand{\relcgd}{$REC[\gamma,\delta]$ }

\newcommand{\bX}{\mathbf{X}}

\newcommand{\bx}{\mathbf{x}}
\newcommand{\bY}{\mathbf{Y}}
\newcommand{\bZ}{\mathbf{Z}}
\newcommand{\by}{\mathbf{y}}

\newcommand{\tZ}{\td{Z}}
\newcommand{\hY}{\hat{Y}}

\newcommand{\btdZ}{\mathbf{\td{Z}}}

\newcommand{\bbP}{\mathbb{P}}

\newcommand{\cX}{{\cal X}}
\newcommand{\cY}{{\cal Y}}
\newcommand{\cZ}{{\cal Z}}
\newcommand{\cS}{{\cal S}}

\newcommand{\cL}{{\cal L}}

\newcommand{\cP}{{\cal P}}
\newcommand{\cA}{{\cal A}}

\newcommand{\cK}{{\cal K}}

\newcommand{\view}{V}

\newcommand{\td}[1]{\tilde{#1}}

\newcommand{\rx}{X}

\newcommand{\removed}[1]{}

\newcommand{\ajb}[1]{\textcolor{black}{{\bf}#1}}

\makeatletter
\patchcmd{\@IEEEeqnarray}{\relax}{\relax\intertext@}{}{}
\makeatother

\title{On Reverse Elastic Channels and the Asymmetry of  Commitment Capacity under Channel Elasticity}

\author{
\IEEEauthorblockN{Amitalok J. Budkuley, Pranav Joshi, Manideep Mamindlapally and Anuj Kumar Yadav}\\
\IEEEauthorblockA{%
}\thanks{\ajb{A short version of this work has been accepted for publication to the IEEE Information Theory Workshop (ITW) 2021, Kanazawa, Japan. This extended version includes detailed proofs, results and discussions.} }
\thanks{A. J. Budkuley, P. Joshi and M. Mamindlapally are with the Department of Electronics and Electrical Communication Engineering, Indian Institute of Technology Kharagpur, West Bengal, India (emails: \texttt{amitalok@ece.iitkgp.ac.in, pranavjoshi@iitkgp.ac.in, manideepyx@iitkgp.ac.in}).  A. K. Yadav is with the Department of Electrical Engineering, Indian Institute of Technology Patna, Bihar, India (email: \texttt{1801ee69@iitp.ac.in}). This work was partially supported by a grant from ISIRD, IIT Kharagpur. The author order is alphabetic. }

}
\begin{document}
	\maketitle 
\thispagestyle{empty}
\begin{abstract}

Commitment is an important cryptographic primitive. It is well known that noisy channels are a promising resource to realize commitment in an \emph{information-theoretically} secure manner. However, oftentimes, channel behaviour may be poorly characterized thereby limiting the commitment throughput and/or degrading the security guarantees;  particularly problematic is when a dishonest party, unbeknown to the honest one,  can maliciously alter the channel characteristics. Reverse elastic channels (RECs) are an interesting class of such \emph{unreliable} channels, where only a  dishonest \emph{committer}, say Alice, can maliciously alter the channel. RECs have attracted recent interest in the study of several cryptographic primitives.

Our principal contribution is the REC commitment capacity characterization; this proves a recent related conjecture. A key result is our tight converse which analyses a specific cheating strategy by Alice. RECs are closely related to the classic unfair noisy channels (UNCs); elastic channels (ECs), where only a dishonest \emph{receiver} Bob can alter the channel, are similarly related. In stark contrast to UNCs, both RECs and ECs always exhibit positive commitment throughput for all non-trivial parameters. Interestingly, our results show that  channels with exclusive \emph{one-sided elasticity} for dishonest parties, exhibit a fundamental \emph{asymmetry} where, a committer with one-sided elasticity has a  more debilitating effect on the commitment throughput than a receiver. 

\end{abstract}

\section{Introduction}\label{sec:introduction}

\input{intro}

\section{Notation and Preliminaries}\label{sec:not:pre}
\input{subsec_notation}
\section{System Model and Problem Description}\label{sec:system:model}
\input{subsec_problem_setup}

\section{Our Main Results}\label{sec:main:results}
\input{main_results}

\section{Converse}\label{sec:proofs:converse}
\input{subsec_converse}

\section{Achievability}\label{sec:proofs:achieve}
\input{subsec_achievability}

\section{Concluding  Remarks and Discussion}\label{sec:conclusion}
\input{conclusion}

\appendix
\input{appendix.tex}

\section*{Acknowledgement}
\ajb{The authors would like to acknowledge several interesting discussions with Manoj Mishra (NISER, HBNI, Bhubaneshwar) which helped us progress in this work.}

\bibliographystyle{IEEEtran}
\bibliography{IEEEabrv,References,refs_AKY}

\end{document}

%% file: intro.tex
Imagine playing a game of rock-paper-scissors, albeit in this time of social distancing. A fundamental conundrum is the following: how does one simulate and verify an instance of \emph{simultaneous play}, an intrinsic feature of this game, among two parties who are fundamentally distrustful and not collocated? 

In essence, each player seeks the following two guarantees vis-\`{a}-vis their opponent: the player, say Alice, can \emph{commit} to her move under the guarantee that her move remains hidden until she chooses to reveal it to the other player, say Bob. Secondly, when revealed, Bob is able to detect precisely whether Alice cheats on her choice. Such a two-phase \emph{commitment} protocol, comprising \emph{commit} followed by \emph{reveal} phases offers exactly the functionality we seek. \footnote{\label{foot:game}\ajb{Our approach in this work follows the \emph{game-based} security paradigm; this differs from an alternate \emph{simulation-based} paradigm (cf.~\cite{canetti2001universally,goldreich2009foundations}). Note however, that for simulators with no computational limitations, the game-based security notion coincides with the simulation-based security notion.}    } In fact, commitment protocols appear as  crucial cryptographic primitives  in several practical applications like sealed-bid auctions~\cite{nojoumian2010unconditionally}, coin flipping~\cite{naor1991bit}, zero knowledge proofs~\cite{Brassard1988}, contract signing~\cite{contract_sign} and secure multiparty computation~\cite{smc1}. 

It is well known that noiseless communication between parties precludes \emph{information-theoretically secure} commitment.\footnote{Blum~\cite{blum}, however, showed that commitment  is possible over one-way noiseless channels when parties are computationally bounded. } Wyner's seminal work~\cite{wyner1} on the \emph{wiretap channel} brought the focus on noisy channels as a resource for realizing information-theoretic security. Commitment (along with a closely related problem called the \emph{oblivious transfer}) has since been widely studied over noisy channels~\cite{crepeau1,crepeau2}. Winter \emph{et al.} characterized the maximum throughput or \emph{commitment capacity} over general discrete memoryless channels (DMCs)~\cite{winter2003commitment}. This was extended to DMCs under fairly general inputs costs in a recent work~\cite{mymb}. Computationally efficient schemes over DMCs have also been studied~\cite{imai}. Commitment has also been explored over continuous channels~\cite{nascimento-barros-t-it2008}, compound channels~\cite{ymbm} as well as quantum channels~\cite{lo1997quantum}. 

Unlike aforementioned works on \emph{fully characterized} noisy channel, the focus of this work is commitment over \emph{unreliable} noisy channels, in particular, the \emph{reverse elastic channel (REC).}

\indent\textbf{Unreliable Channels:} 
Oftentimes access to a noisy channel may be under incomplete knowledge of the channel law; in such cases, direct use of existing commitment schemes for fixed channel may severely degrade the security guarantees for the two parties. Damg{\aa}rd \emph{et al.}~\cite{damgaard1999possibility} initiated a systematic study of unreliable channels and proposed the \emph{unfair noisy channels (UNCs)}. 
%
\begin{definition}[Unfair noisy channel (UNC)]\label{def:UNC}
An unfair noisy channel (UNC) with parameters  $0<\gamma<\delta<1/2$, also called UNC$[\gamma,\delta]$, is a noisy BSC where (i) honest parties communicate over a $BSC(s)$, where $s\in\cS=[\gamma,\delta]$ and unknown to them, (ii) any dishonest party can privately set $s$ to any value in $\cS$.\footnote{Along with the noiseless binary symmetric channel (BSC), it can be shown that commitment is impossible over a BSC($1/2$). Hence, $\gamma$, $\delta$ are so chosen.}
\end{definition} 
%
Unlike fixed channels (for instance, classic BSC),  UNCs introduce an asymmetry in the capabilities of a party vis-\`{a}-vis channel awareness and control when said party is honest and when it is dishonest. 
%
	Interestingly, commitment was shown to be impossible over UNCs when $\delta \geq 2\gamma(1-\gamma)$ in~\cite{damgaard1999possibility}; the commitment capacity, however, was only recently characterized (the converse in~\cite[remark on pg.4]{crepeau2020commitment} was presented for commitment schemes under some restrictions) in~\cite{crepeau2020commitment} and shown to be $H(\gamma)-H\left(\frac{\delta-\gamma}{1-2\gamma}\right)$. While both parties Alice and Bob have identical capabilities (when honest/dishonest) in a UNCs, more recent works have studied models when capabilities are fully skewed or \emph{one-sided}. In~\cite{khurana2016secure,elastic_2}, the elastic channel (EC) is studied. 
\begin{definition}[Elastic channel (EC)]\label{def:EC}
An elastic channel (EC)  with parameters $0<\gamma<\delta<1/2$, also called EC$[\gamma,\delta]$, is a noisy BSC  where (i) honest parties communicate over a classic $BSC(\delta)$, (ii) only a dishonest Bob can privately set the crossover probability to any value $s$ in $\cS=[\gamma,\delta]$. 
\end{definition} 
In RECs, which are the focus of this work, the capabilities of Alice and Bob are however reversed.
\begin{definition}[Reverse elastic channel (REC)]\label{def:REC}
A reverse elastic channel (REC)  with parameters $0<\gamma<\delta<1/2$, also called REC$[\gamma,\delta]$, is a noisy BSC where (i) honest parties communicate over a classic $BSC(\delta)$, (ii) only a dishonest Alice can privately set the crossover probability to any value $s$ in $\cS=[\gamma,\delta]$. 
\end{definition} 
As can be seen, unlike in UNCs, the capabilities of Alice and Bob, when dishonest, differ significantly in  ECs and RECs. Essentially, in such channels with \emph{one-sided elasticity,} the REC (resp. EC)  allows exclusive individual channel control to a dishonest committer Alice (resp. receiver Bob), unbeknown to the receiver Bob (resp. committer Alice).  

In~\cite{crepeau2020commitment}, the authors also presented the commitment capacity of the elastic channel EC$[\gamma,\delta]$ and showed it to be $H(\gamma).$ However, a conjecture without proof was made for the capacity of the REC$[\gamma,\delta].$ In this work, we show the conjecture to be true and present the capacity characterization of the REC$[\gamma,\delta]$. 

\indent\textbf{Contributions:}
The following are our principal contributions:\\
\noindent$\bullet$ We completely characterize the commitment capacity $\mathbb{C}_{REC}$ of an REC$[\gamma,\delta]$ (cf. Theorem~\ref{thm:rec:capacity}); we show that $\mathbb{C}_{REC}=H(\delta)-H\left(\frac{\delta-\gamma}{1-2\gamma}\right)$.\\
\noindent$\bullet$ We present a novel converse where we analyse a specific cheating strategy by Alice (cf. Sec.~\ref{sec:proofs:converse}); the analysis is inspired by the converse for UNCs (cf.~\cite{crepeau2020commitment}) but differs significantly in getting the optimal rate bound. Crucially, unlike in~\cite{crepeau2020commitment} where the authors restrict their converse to commitment schemes which need to satisfy a special  Markov chain, we prove our converse under no such limitation, and with complete generality (see the discussion after Theorem~\ref{thm:rec:capacity} for a detailed discussion). We also present an optimum achievability scheme (cf. Sec.~\ref{sec:proofs:achieve}).\\
\noindent$\bullet$ Our results reveal the following fundamental asymmetry: under identical \emph{one-sided elasticity,} a malicious committer degrades the commitment throughput more than a malicious receiver. We leverage this insight to propose a significantly generalized framework of elastic channels with \emph{two-sided elasticity}. We then present a conjecture on the commitment capacity of its \emph{symmetric} channel instance, viz., the symmetric two-sided elastic channel. 

\indent\textbf{Organization of paper:} The rest of the paper is organized as follows: In Section~\ref{sec:not:pre},
we present the notation and preliminaries used in this work. In Section~\ref{sec:system:model}, we describe the problem setup and state the problem. In Section~\ref{sec:main:results}, we present our commitment capacity characterization for REC$[\gamma,\delta]$. In Sections~\ref{sec:proofs:converse} and~\ref{sec:proofs:achieve}, we present the  proof details for the converse and achievability respectively. We  make concluding remarks in Section~\ref{sec:conclusion}, followed by the appendices which include supporting proofs in detail.

%% file: subsec_notation.tex
%
We denote random variables by upper case letters (eg. $X$), the values they take by lower case letters (eg., $x$), and their alphabets by calligraphic letters (eg. $\cX$). Unless stated otherwise, all sets are assumed to be finite. We denote random vectors and the accompanying values they take by boldface letters (e.g., $\bX=(X_1,X_2,\cdots,X_n)$, $\bx=(x_1,x_2,\cdots,x_n)$, resp.).
For any natural number $a\in\mathbb{N}$, let $[a]:=\{1,2,\cdots, a\}$. 
We denote the Hamming distance between two vectors, say $\vec{x},\vec{x}'\in\cX^n$ by $d_H(\vec{x},\vec{x}').$
%
Let $P_X$ denote the distribution of $X\in\cX$; $\cP(\cX)$ denotes the simplex of probability distributions on set $\cX$. Distributions for multiple random variables are similarly defined. 
%
Let $\mathbb{P}(A)$ denote the probability of event $A$. Deterministic and random functions will be denoted by lower case letters (eg. $f$) and by upper case letters (e.g., $F$) respectively. Let $X\sim \text{Bernoulli}(p)$ denote a Bernoulli random variable $X$ with parameter $p\in[0,1]$. Let $p*q:=p(1-q)+(1-p)q$, where $p,q\in[0,1].$ Given $P_\rx,Q_\rx \in \cP(\cX)$, 
let $||P_{\rx}-Q_\rx||$ denote the statistical (or variational) distance  between $P_\rx$ and $Q_{\rx}$.

Next, we define some classic information measures (cf.~\cite{bloch,csiszar-korner-book2011}). Let random variables $X,Y\in\cX\times\cY$, where $(X,Y)\sim P_{X,Y}$. Then, $H(X)$ and $I(X;Y)$  denote the (Shannon) entropy of $X$ and mutual information of the pair $(X,Y)$ resp.. The \emph{min-entropy} of $X$ is denoted by $H_{\infty}(X):=\min_{x\in\cX} \left(-\log(P_{X}(x))\right)$; the conditional version is given by $H_{\infty}(X|Y):=\min_{y} H_{\infty}(X|Y=y).$ For $\epsilon\in[0,1)$, the \emph{$\epsilon$-smooth min entropy} and its conditional version is given by:
$H_{\infty}^\epsilon(X):= \max_{X':||P_{X'}-P_{X}||\leq\hspace{1mm}\epsilon} H_{\infty}(X')$ and 
$H_{\infty}^\epsilon(X|Y):= \max_{X',Y':||P_{X',Y'}-P_{X,Y}||\leq\hspace{1mm}\epsilon} H_{\infty}(X'|Y')$ respectively. We also need universal hash functions and strong randomness extractors for our commitment scheme; we describe them next.
\begin{definition}[$\xi$-Univeral hash functions~\cite{carter_universal_1979}] 
	Let $\mathcal{H}$ be a class of functions from $\cX$ to $\cY$. $\mathcal{H}$ is said to be $\xi-$universal hash function, where $\xi\in\mathbb{N}$, if when $h\in\mathcal{H}$ is chosen uniformly at random, then $(h(x_1),h(x_2),...h(x_{\xi}))$ is uniformly distributed over $\cY^{\xi}$, $\forall x_1,x_2,...x_{\xi} \in \cX$.
\end{definition}
\begin{definition}[Strong randomness extractors~\cite{nisan1996randomness,glh}]
	 A probabilistic polynomial time function of the form \text{Ext}: $\{0,1\}^n \times \{0,1\}^d \to \{0,1\}^m$ is an  $(n,k,m,\epsilon$)-strong extractor if for every probability distribution $P_{Z}$ on $\cZ=\{0,1\}^n$, and $H_{\infty}(Z)\geq k$, for random variables $D$ (called 'seed') and $M$, distributed uniformly in $\{0,1\}^d$ and $\{0,1\}^m$ respectively, we have $||P_{Ext(Z;D),D}- P_{M,D}|| \leq \epsilon$.
\end{definition}

%% file: subsec_problem_setup.tex
\begin{figure}[!ht]
  \begin{center}
		\includegraphics[trim=1cm 8cm 0cm 1cm, scale=0.4]{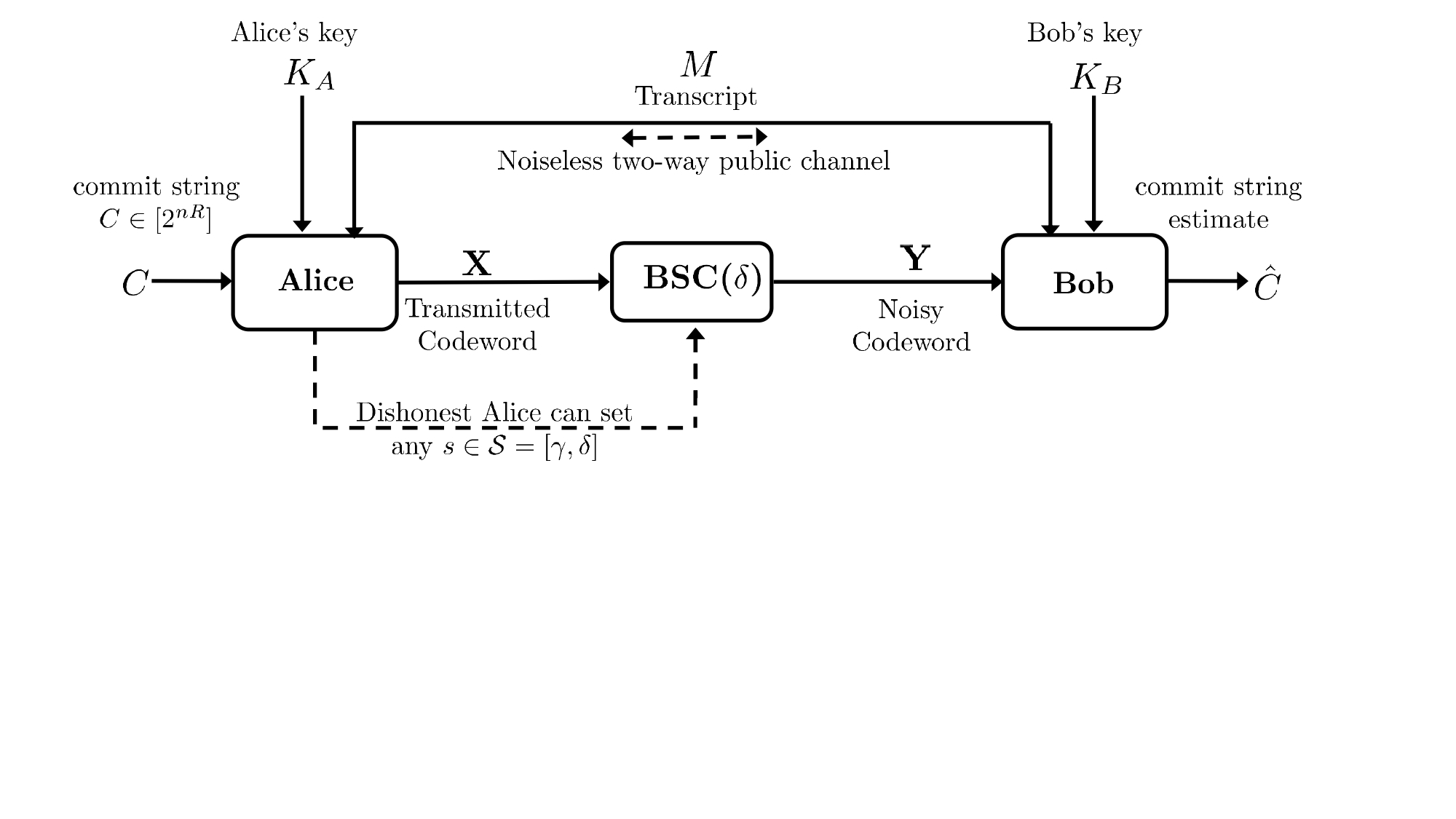}
    \caption{The problem setup: commitment over an REC$[\gamma,\delta]$ }
    \label{fig:main:setup}
  \end{center}
\end{figure}

Fig. \ref{fig:main:setup} depicts the commitment problem setup comprising  two mutually distrustful parties, the \emph{committer} Alice and the \emph{receiver} Bob.  Alice seeks to commit to a bit string $C \in [2^{nR}]$, where rate $R>0$ is specified later. They have access to a  one-way (Alice-to-Bob) noisy REC$[\gamma,\delta]$, where $0<\gamma<\delta<1/2$ (cf. Definition~\ref{def:REC}). Apart from the REC$[\gamma,\delta]$, Alice and Bob can also  communicate over a two-way noiseless authenticated public channel. Alice makes  $n$ uses of REC$[\gamma,\delta]$. Let $\bX$ denote her channel input; Bob receives its  noisy version $\bY$. 
%
%
\ajb{Both Alice and Bob can privately randomize. Alice's key $K_A \in \cK_A$ and Bob's key $K_B \in \cK_B$ are independent and generated privately via random experiments; these model the randomness in Alice's and Bob's actions and/or transmissions in the protocol.} At any point in time, any message transmitted by individual parties can depend causally on the information available to them. 

\label{page:protocol}
We now define a \emph{commitment protocol} over REC$[\gamma,\delta]$.
\begin{definition}[$(n,R)$-commitment protocol]
An $(n,R)$-commitment protocol $\mathscr{P}$ is a procedure of message exchange between Alice and Bob over two phases, comprising commit phase followed by reveal phase, with the aim of committing over a uniformly random string $C \in [2^{nR}]$. Here  $R \in [0,1]$ is the \emph{rate} of the $(n,R)$-commitment protocol.\footnote{\label{foot:rate}
\ajb{Similar to other works, we assume that Alice and Bob have \emph{prior} access to a two-way noiseless link and  define the \emph{rate} of a commitment protocol as the ratio of the length of the committed string  to the number of invocations of the one-way noisy channel, which in this case is an REC, from Alice to Bob. However, it is pertinent to note that there exist alternate notions of rate. For example, one could additionally invoke the REC to realize  a reliable two-way communication link, and then amortize the size of the commitment string over the overall number of REC invocations in the protocol. We do not explore this notion of rate in our work.  
}
}\label{page:rate}

\indent $(a)$ \emph{Commit phase:} Given $C\in[2^{nR}]$, Alice sends a   vector $\bX$ over $n$ uses of the the \relcgd; Bob, in response  receives $\bY$. In between the transmissions over the \relcgd, Alice and Bob also exchange messages  over the noiseless two-way public channel available to them;\footnote{The messages exchanged over the two-way noiseless channel may be arbitrarily large but finite in size.} the entire \emph{transcript} of the messages is denoted by $M$. Alice and Bob's views, denoted by $V_A$ and $V_B$ respectively, comprise the collection of random variables and/or vectors known to them at the end of the commit phase. In particular, we have $V_A = (C,\bX,K_A, M)$, and $V_B = (\bY, K_B, M)$.

\indent $(b)$ \emph{Reveal phase:} Alice and Bob only communicate over the public channel. Alice announces to Bob the pair comprising $\bar{c} \in [2^{nR}]$ and a vector $\bar{\bx}\in \{0,1\}^n$. Upon receiving $(\bar{c},\bar{\bx})$, Bob performs a test $T(\bar{c},\bar{\bx}, V_B) \in \{0,1\}$, and based on the outcome of the test, accepts  $\bar{c}$ as the commit string if the test passes ($T=1$) and rejects if the test fails ($T=0$).
\end{definition}

We now define for this $(n,R)$-commitment protocol $\mathscr{P},$ the following key parameters in the context of the REC$[\gamma,\delta]$:

\begin{definition}[$\epsilon$-sound]
Protocol $\mathscr{P}$ is said to be $\epsilon$-sound if for an honest Alice and  an  honest Bob,
\begin{equation}
   \ajb{\max_{c\in[2^{nR}]} \mathbb{P}\big( T(c,\bX,V_B)=0\big)\leq \epsilon }   \label{eq:sound}
\end{equation}
\end{definition}

\begin{definition}[$\epsilon$-concealing]
Protocol $\mathscr{P}$ is said to be $\epsilon$-concealing if for an \emph{honest} Alice, under any strategy of Bob,
\begin{equation*}
    I(C;\view_B)\leq \epsilon.\label{eq:concealing}
\end{equation*}
\end{definition}

\begin{definition}[$\epsilon$-binding]
Protocol $\mathscr{P}$ is said to be $\epsilon$-binding if for an honest Bob, and any strategy of Alice
\begin{equation*}
    \max_{s\in[\gamma, \delta]} \mathbb{P}\Big(T(\bar{c},\bar{\bx},V_B)=1 \quad\& \quad T(\hat{c},\hat{\bx},V_B)=1\Big| S=s\Big)\leq \epsilon\label{eq:binding}
\end{equation*}
for any two pairs $(\bar{c},\bar{\bx})$, $(\hat{c},\hat{\bx})$, $\bar{c}\neq \hat{c}$ and $\bar{\bx},\hat{\bx}\in\{0,1\}^n$.
\end{definition}

A rate $R\in[0,1]$ is said to be \emph{achievable} if for every $\epsilon>0$, there exists for every $n$ sufficient large, an  $(n,R)$-commitment protocol  which is $\epsilon$- sound, $\epsilon$-concealing and $\epsilon$-binding. The supremum of all achievable rates is defined as the \emph{commitment capacity} of the REC$[\gamma,\delta]$, denoted by $\mathbb{C}_{REC}.$

%% file: main_results.tex
The principal contribution  of this work is the commitment capacity characterization of the REC$[\gamma,\delta].$ 
\begin{theorem}[REC commitment capacity]\label{thm:rec:capacity}
\ajb{The commitment capacity of the REC$[\gamma,\delta]$, where $0<\gamma<\delta<1/2$, is 
\begin{equation}\label{eq:rec:capacity}
\mathbb{C}_{REC}=H(\delta)-H\left(\kappa\right),
\end{equation}
where $\kappa:=\frac{\delta-\gamma}{1-2\gamma}$ and $\delta=\gamma*\kappa.$}
\end{theorem}
%

Our result proves the conjecture stated in~\cite{crepeau2020commitment} on RECs. 
A key contribution of our work is the matching rate upper bound (see Section~\ref{sec:proofs:converse}). Although our converse analysis is inspired by the approach in~\cite{crepeau2020commitment} for UNCs, it has some novel differences. Crucially, we prove our converse under complete generality, unlike the one for UNCs in~~\cite{crepeau2020commitment}. In that work, the authors  impose a condition where the Markov chain $M\leftrightarrow \bY\leftrightarrow \bX$ holds; this is restrictive and commitment protocols in general need not satisfy such a condition (this limitation is also pointed out in~\cite{crepeau2020commitment}). Additionally, for the specific cheating strategy of Alice, the authors leverage a degraded channel structure over the UNC; such a structure is not available over the REC which necessitates a different approach. See Sec.~\ref{sec:proofs:converse} for the detailed converse proof. 

Our achievability commitment protocol follows Damg{\aa}rd \emph{et al.'s} construction~\cite{damgaard1999possibility}. \ajb{In particular, our presentation is inspired by~\cite{crepeau2020commitment}; however, we analyse a soundness criterion where \emph{every} commit string $c\in[2^{nR}]$ is accepted with a probability of at least $1-\epsilon$. This is  \emph{stronger}  than the corresponding criterion in~\cite{crepeau2020commitment} where on average (over $C\in[2^{nR}]$) soundness is guaranteed. \footnote{\ajb{It is known that for some problems such a change in the criterion can lead to different notions of `capacity' (see, for instance,~\cite{lapidoth-narayan-it1998}). However, commitment capacity remains the same for both \emph{average} and \emph{maximal} soundness criteria here.}}   We refine the choice of the protocol parameters for the given REC and analyse soundness, concealment and bindingness (see Section~\ref{sec:proofs:achieve}) of the protocol.  An interesting consequence of this work  is that even when the malicious party is \emph{adaptively} allowed to set potentially different values $s_i\in[\gamma,\delta]$ for $i\in[n],$ there is essentially no benefit to the said party as no further commitment rate degradation is possible (this is also seen in UNCs; see~\cite{damgaard1999possibility} for instance). }
\begin{figure}[!ht]
  \begin{center}
    \includegraphics[clip,trim=3cm 9cm 0cm 8cm, scale=0.65]{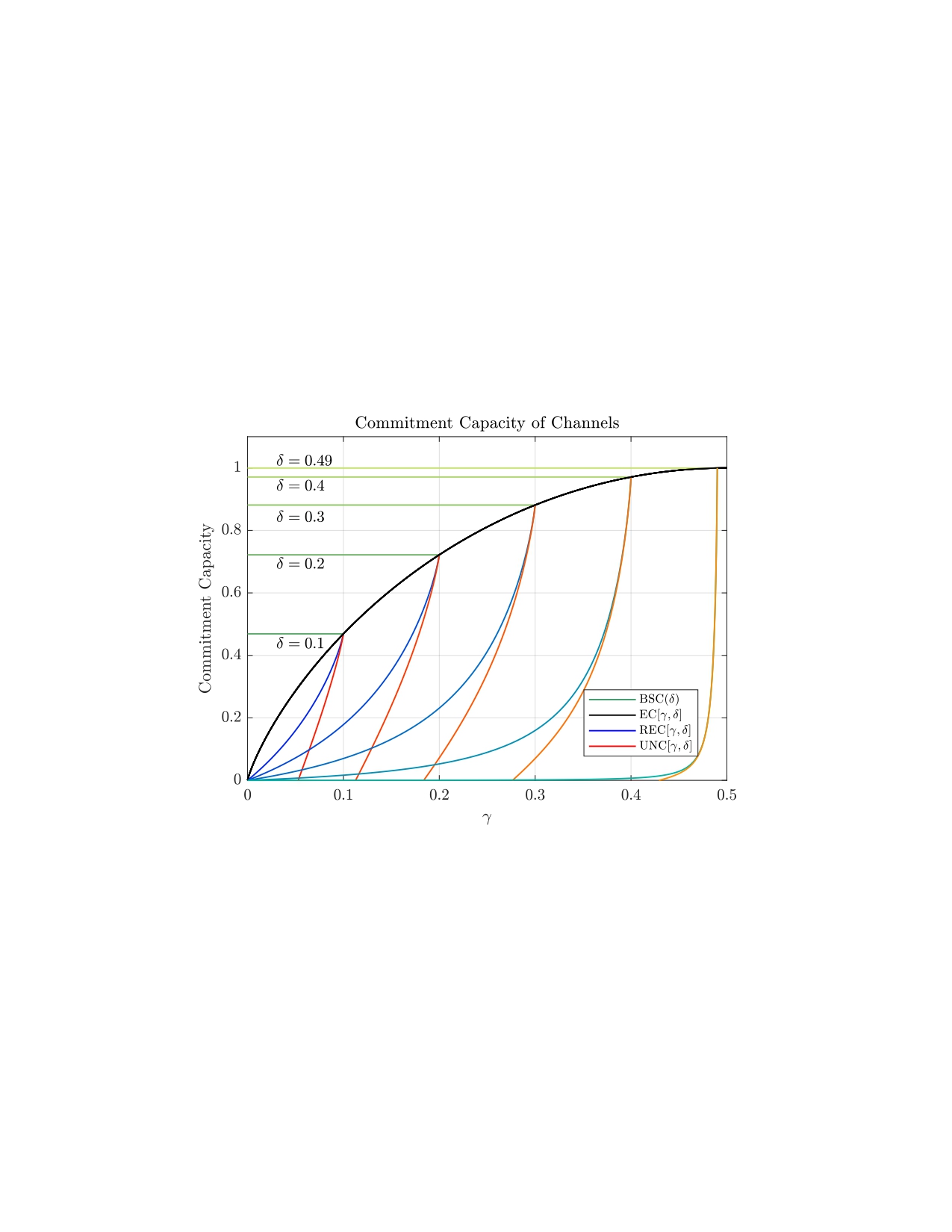}
    \caption{Variation of commitment capacities (w.r.t. $\gamma$) for different channels. Curves are presented for different values of $\delta\in(0,1/2).$ }
    \label{fig:capacity:curves}
  \end{center}
\end{figure}

From our result in Theorem~\ref{thm:rec:capacity} and the corresponding results for ECs and UNCs (cf.~\cite{crepeau2020commitment}), we can establish that $\mathbb{C}_{EC}> \mathbb{C}_{REC}>\mathbb{C}_{UNC}$ for any  specified $\gamma,\delta$ values. Refer Fig.~\ref{fig:capacity:curves} where we plot the capacities of these \emph{unreliable} channels along with the BSC($\delta$). 
\begin{remark}[Positive commitment throughput]\label{rem:pos}
Unlike UNC$[\gamma,\delta]$ which may have zero commitment capacity (this occurs when $\delta\geq \gamma*\gamma:=2\gamma(1-\gamma)$, see~\cite{crepeau2020commitment}), an REC$[\gamma,\delta]$ always exhibits  positive commitment capacity for the specified range of parameters. Note that the same is true for an  EC$[\gamma,\delta]$ whose capacity is $\mathbb{C}_{EC}=H(\gamma)>0$~\cite{crepeau2020commitment}. 
\end{remark}

The following is a key takeaway from this work: commitment throughput over RECs is strictly lower than that over ECs (under identical $\gamma,\delta$ parameters) when parties can  malicious alter the channel characteristics. This fact reveals an interesting \emph{asymmetry} in commitment over such unreliable channels with \emph{one-sided elasticity}, i.e., channels which afford elasticity (i.e., capability to alter the channel) to exactly one of the dishonest parties exclusively. Essentially, a dishonest \emph{committer} Alice \emph{always} degrades the commitment throughput more than a dishonest \emph{receiver} Bob.  \ajb{This is in stark contrast to the \emph{symmetric} scenario under \emph{honest-but-curious} parties which  lack malicious channel control; the REC (as well as EC) essentially defaults to a classic BSC($\delta$) here. For such honest-but-curious adversaries, RECs and ECs  offer identical commitment throughput.}\label{page:curious}

\begin{figure}[!ht]
  \begin{center}
    \includegraphics[clip, trim=2cm 8cm 0cm 5cm, scale=0.5]{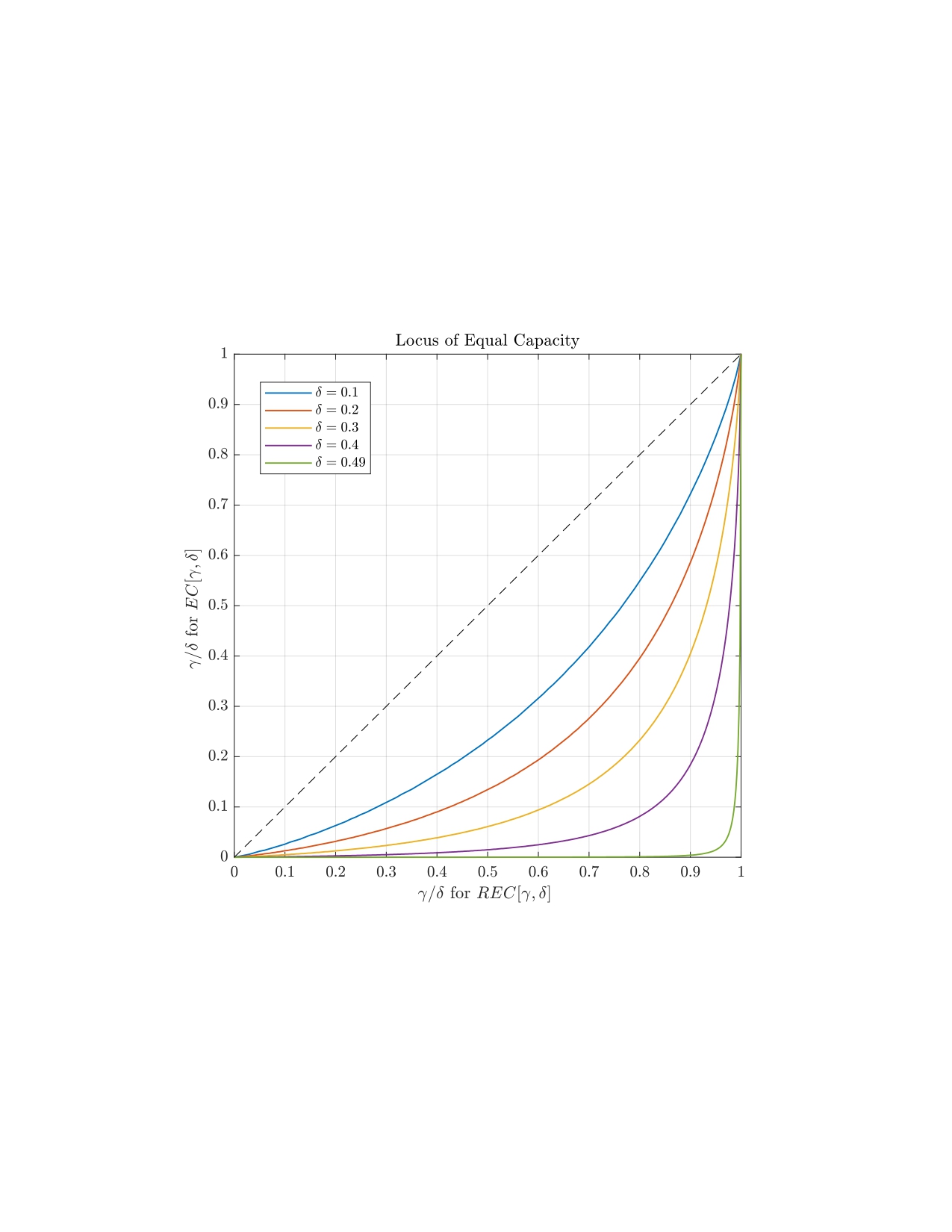}
    \caption{The EC$[\gamma,\delta]$ versus REC$[\gamma,\delta]$ commitment capacity contour plotted when those capacities are identical. Curves are presented for different values of $\delta\in(0,1/2).$  }
    \label{fig:contour:capacity}
  \end{center}
\end{figure}

Fig.~\ref{fig:contour:capacity} illustrates the asymmetry in the commitment capacity for the RECs and the ECs more succinctly; in Fig.~\ref{fig:contour:capacity} we present the joint `equal-capacity' contours for RECs and ECs. As can be seen in Fig.~\ref{fig:contour:capacity}, for a fixed $\delta\in(0,1/2)$, a dishonest receiver in EC$[\gamma,\delta]$ requires considerably `larger' receiver-side elasticity, characterized by a lower $\gamma$ (the axes plot a normalized value of $\gamma$ w.r.t. $\delta$), to effect the same degradation of the commitment throughput than a dishonest committer in an REC$[\gamma,\delta]$. Furthermore, as $\delta$ increases, one can observe that the skew in the asymmetry, which essentially characterizes the committer-receiver `mismatch' in `elastic-capabilities', is more pronounced.

Seen from another perspective, for a fixed $\delta\in(0,1/2)$, the \emph{gap} in the commitment capacity $\Omega_{\delta}(\gamma):=\mathbb{C}_{EC}-\mathbb{C}_{REC}$ is strictly positive (note that $0<\gamma<\delta<1/2$), though it is \emph{not} a constant (see Fig.~\ref{fig:capacity:curves}). Furthermore, this gap $\Omega_{\delta}(\cdot)$ increases as $\delta$ increases in the range $(0,1/2);$ it can be  shown that  $\Omega_{\delta}(\gamma)$ is concave in $\gamma$ (for fixed $\delta$), and $\Omega_{\delta}(\gamma)$ is maximized when $\gamma\otimes\gamma=\delta$, i.e., for a unique optimizer $\gamma^*(\delta)=\frac{1-\sqrt{1-2\delta}}{2}.$  It is pertinent to note that $\gamma^*(\delta)$ is exactly the value for which the corresponding UNC$[\gamma^*,\delta]$ has zero capacity. 
%
%

%% file: subsec_converse.tex
Consider a sequence of protocols $\{\mathscr{P}\}_{n\geq 1}$. Here every protocol $\mathscr{P}_n$ is $\epsilon_n$-sound, $\epsilon_n$-concealing and $\epsilon_n$-binding, where $\epsilon_n\rightarrow 0$ as $n\rightarrow \infty.$

\textbf{Alice's `achievable' strategy:} We analyse the following specific `dishonest' strategy by Alice, feasible for the REC$[\gamma,\delta]$:\footnote{Note that fixing such a strategy gives us an upper bound on rate; in our case, this bound will prove tight.} Alice sets the REC$[\gamma,\delta]$ to a BSC($s$), $s\in[\gamma,\delta]$. Correspondingly, she also sets up a `private' BSC($\kappa_s$), where $\kappa_s:=\frac{\delta-s}{1-2s}\geq 0$; we denote the output of this private BSC($\kappa_s$) as $Z$ (the dependence on $s\in\cS$ is implicit). Note that essentially the channel from $Z$ to $Y$ (via $X$) is always\footnote{This follows from noting that  $\kappa_s\otimes s=\kappa_s(1-s)+(1-\kappa)s$ equals $\delta$ for every $s\in[\gamma,\delta]$.} a BSC($\delta$). We show later that Alice's rate-minimizing choice $s^*$ equals $\gamma$ which results in the tight rate bound we seek.\footnote{Another rate bound, for instance, can be obtained by assessing the case when Alice is `honest', and sets $s=\delta$. However, it is not hard to argue that the resulting rate bound $R\leq H(\delta)$ will  only be `weak'.}

Such a cheating Alice sends $\bX$ over the BSC($s$) to Bob, and privately generates $\bZ$ by passing $\bX$ through the private channel BSC($\kappa_s$); given that the pair $(\bZ,\bY)$ are `compatible' over the BSC($\delta$), we have $\mathbb{P}\big( T(C,\bZ,V_B)=0\big)\leq \epsilon_n,$ where $T$ is Bob's test. Let us denote $\td{Z}:=(Y,Z)$, and let $\td{\bZ}:=(\bY,\bZ)$.  

We now state two useful lemmas used later in our analysis. 
\begin{lemma}\label{lem:fano:bindsound}
For every $\mathscr{P}_n$ which is $\epsilon_n$-sound and $\epsilon_n$-binding, $H(C|\bZ,\bY,K_B, M)\leq n\epsilon'_n$, where $\epsilon'_n(\epsilon_n)\rightarrow 0$ as $\epsilon_n\rightarrow 0$.
\end{lemma}
The proof of this lemma appears in Appendix~\ref{app:fano:lemma}.  Note that our converse holds in full generality (see proof details later); this is quite unlike in the converse for UNCs~\cite{crepeau2020commitment} where the authors require that commitment protocols  satisfy the Markov chain $M\leftrightarrow \bY\leftrightarrow \bX,$ thereby restricting the validity of the rate upper bound to those protocols only. 

Let $\td{\bZ}^{i}:=(\tZ_1, \tZ_2,\cdots,\tZ_{i})$ and let $\hat{\bY}^{i}:=(Y_{i}, Y_{i+1}, \cdots, Y_n)$. The following lemma is stated without proof (the proof follows directly from~\cite{csiszar1978broadcast}).
\begin{lemma}[\cite{csiszar1978broadcast}]\label{lem:czissar:seperation}
%
Let $W:=(K_B,M)$. Then, 
\begin{IEEEeqnarray*}{rCl}
&I(C&;\btdZ|W) - I(C;\bY|W) = \sum_{i=1}^{n} [I(C;\tZ_i | W, \td{\bZ}^{i-1},\hat{\bY}^{i+1}) \notag\\ 
&&\hspace{40mm}- I(C;Y_i |W, \td{\bZ}^{i-1}, \hat{\bY}^{i+1})].\label{eq:czissar:seperation}
\end{IEEEeqnarray*}
\end{lemma}

We now bound the rate $R$ of the commitment protocol $\mathscr{P}_n$:
\begin{IEEEeqnarray}{rCl}
nR&=& H(C)  \notag\\
&\stackrel{}{=}& H(C|V_B) + I(C;V_B)  \notag\\
&\stackrel{(a)}{\leq}& H(C|\bY, K_B, M) + \epsilon_n  \notag\\
&\stackrel{(b)}{=}& H(C|\bY, K_B, M) - H(C|\bY,\bZ, K_B, M)\notag\\
&&\hspace{5mm} + H(C|\bY,\bZ, K_B M) + \epsilon  \notag\\
&\stackrel{(c)}{\leq}& H(C|\bY, K_B, M) - H(C|\bY,\bZ, K_B, M) + n\epsilon'_n + \epsilon_n  \notag\\
&\stackrel{(d)}{=}& I(C;\bY, \bZ | K_B, M) - I(C;\bY |K_B, M) + n\epsilon'_n + \epsilon_n  \notag\\
&\stackrel{(e)}{=}& I(C;\btdZ | K_B, M) - I(C;\bY |K_B, M) + n\epsilon'_n + \epsilon_n  \notag \\
&\stackrel{(f)}{\leq}& \sum_{i=1}^{n} [I(C;\tZ_i | K_B, M, \tZ^{i-1}, \hY^{i+1} )  \nonumber\\ 
&&\hspace{5mm} - I(C;Y_i |  K_B, M, \tZ^{i-1}, \hY^{i+1}) ] + n\epsilon'_n + \epsilon_n  \label{eq:rate:1}
\end{IEEEeqnarray}
where we have
\begin{enumerate}[(a)]
    \item as $\mathscr{P}_n$ is $\epsilon_n$-concealing, and from the definition of $V_B$. 
    \item by adding and subtracting $H(C|\bY,\bZ, K_B, M)$
		\item from Lemma~\ref{lem:fano:bindsound}
    \item by adding and subtracting $H(C|K_B, M)$
    \item from the definition of $\td{\bZ}$
    \item from Lemma~\ref{lem:czissar:seperation}.
\end{enumerate}
To proceed from~\eqref{eq:rate:1}, let us define an independent random variable $L\sim \text{Unif}([n])$. Also, let 
$U:= (K_B, M, \td{\bZ}^{L-1}, \hat{\bY}^{L+1}, L)$
$V:= (U,C).$
Observe that $U$ depends only on $\tZ_i$, $i<L$, and and $Y_j$, $j>L$. Furthermore, $Y_L$ is a trivially degraded version of $\tZ_L = (Y_L, Z_L)$. Thus, we have the following Markov chain: $U \leftrightarrow V \leftrightarrow X \leftrightarrow \tZ \leftrightarrow Y$. 

We now use these facts to simplify~\eqref{eq:rate:1} as follows:
\begin{IEEEeqnarray}{lCl}
R
&\stackrel{(a)}{\leq}&\sum_{i=1}^{n} \mathbb{P}(L=i) [ I(C;\tZ_L | K_B, M ,\tZ^{L-1}, \hY^{L+1}, L=i )\notag\\
&&\hspace{5mm} - I(C;Y_L |  K_B, M, \tZ^{L-1}, \hY^{L+1}, L=i) ] + \epsilon'_n + \td{\epsilon}_n \notag\\
&\stackrel{(b)}{=}& I(C;\tZ|U) - I(C;Y|U) + \epsilon'_n + \td{\epsilon}_n  \notag\\
&\stackrel{(c)}{=}& I(V;\tZ|U) - I(V;Y|U) + \epsilon'_n + \td{\epsilon}_n  \notag\\
&\stackrel{(d)}{=}& I(V;\tZ) - I(U;\tZ) - I(V;Y) - I(U;Y) + \epsilon'_n + \td{\epsilon}_n  \notag\\
&\stackrel{(e)}{=}& I(X;\tZ) - I(X;Y) - [ I(X;\tZ|V) - I(X;Y|V)]\notag\\
&&\hspace{5mm} - [ I(U;\tZ) - I(U;Y)] +\epsilon'_n + \td{\epsilon}_n\nonumber \\
&\stackrel{(f)}{\leq}& I(X;\tZ) - I(X;Y) + \epsilon'_n + \td{\epsilon}_n\label{eq:ratebound:mutualinfo}
\end{IEEEeqnarray}
where we have
\begin{enumerate}[(a)]
    \item from definition of $L$, and letting $\td{\epsilon}_n:=\frac{\epsilon_n}{n}$. 
    \item from noting that $U=(K_B, M, \td{\bZ}^{L-1}, \hat{\bY}^{L+1}, L)$ and letting $X:=X_L$, $Y:=Y_L$ and $\td{Z}:=\tZ_L$.
    \item from noting that $V = (U,C)$.
    \item from the chain rule of mutual information
    \item from the Markov chains  $V \leftrightarrow X \leftrightarrow \tZ$ and $V \leftrightarrow X \leftrightarrow Y$, and non-negativity of the trailing two terms in brackets.
    \item from the Markov chain $X \leftrightarrow \tZ \leftrightarrow Y$ as $Y$ is a degraded version of $\tZ$. 
\end{enumerate}
Note that \eqref{eq:ratebound:mutualinfo} holds $\forall s\in[\gamma,\delta]$.  Letting $n\rightarrow \infty$ and optimizing Alice's choice $s\in[\gamma,\delta]$ (recall her cheating strategy), we have
\begin{align}
R &\leq \min_{s\in [\gamma, \delta]} I(X;YZ) - I(X;Y)\notag\\
  &\stackrel{(a)}{\leq} \max_{P_X}\min_{s\in [\gamma,\delta]} I(X;YZ) - I(X;Y)\notag\\
	&\stackrel{(b)}{=} H(\delta)-H\left(\frac{\delta-\gamma}{1-2\gamma}\right) \notag\\
	&\stackrel{(c)}{=} \mathbb{C}_{REC},\notag
\end{align}
where $(a)$ follows by optimizing the input distribution $P_X$, and $(b)$ follows by optimizing the expression $I(X;YZ) - I(X;Y) = H(X|Y) - H(X|YZ)$ which occurs at
input $X\sim\text{Bernoulli} (1/2)$ and $s^*=\gamma;$ the optimum value equals $H(\delta)-H(\frac{\delta-\gamma}{1-2\gamma}).$ Finally, $(c)$ follows from~\eqref{eq:rec:capacity}.

%% file: subsec_achievability.tex
\noindent Following~\cite{damgaard1999possibility}, our protocol utilizes two rounds of random hash exchange challenges and a strong randomness extractor based on 2-universal hash functions; our presentation is inspired by~\cite{crepeau2020commitment}. The two rounds\footnote{We need two rounds of hash challenge to circumvent a non-trivial rate loss that arises in the single hash challenge due to the \emph{birthday paradox}; see~\cite{crepeau2020commitment} where it is discussed in detail.} of hash challenges essentially \emph{bind} Alice to her choice in the commit phase thereby ensuring Bob's test $T$ can detect any cheating attempt by Alice during the reveal phase. 
The strong randomness extractor extracts a secret key (note that the leftover hash lemma~\cite{glh} allows us to quantify the size of this key). This key is then XOR-ed with the commit string $c$ to realize a \emph{one-time pad} scheme, which conceals the committed string against Bob in the commit phase.

Here are the details of our protocol.  The rate $R:=H(\delta)-H(\kappa)- \beta_3$, where the choice of $\beta_3>0$ is specified later. 
Let $\mathcal{G}_1:=\{g_1:\{0,1\}^n \rightarrow \{0,1\}^{n(H(\kappa)\removed{\max_{s\in\cS \cup \cS_A}H(\frac{\delta-s}{1-2s})} + \beta_1)}\}$ be a $4n$-universal hash family,
where $\kappa:=\frac{\delta-\gamma}{1-2\gamma}$ and $\beta_1>0$ is a small enough constant. 
Let  
$\mathcal{G}_2:=\{g_2:\{0,1\}^n \rightarrow \{0,1\}^{n\beta_2}\}$ be a $2-$universal hash family, where $\beta_2>0$ is a small enough constant.
Let $\mathcal{E}:=\{\text{ext}:\{0,1\}^n \rightarrow \{0,1\}^{nR}\}$ be a $2-$universal hash family, where $\beta_3>0$ is chosen such that $\beta_3 > \beta_1 + \beta_2$.\footnote{Note that  $R$ can be made arbitrarily close to $\mathbb{C}_{REC}.$}

We now describe the commit and reveal phases:\\
\indent $\bullet$\emph{Commit Phase:}  
To commit string $c\in[2^{nR}]$, the protocol proceeds as follows:\\
\noindent (C1). Given $c$, Alice sends $\bX\sim \text{Bernoulli}(1/2)$ independent and identically distributed (i.i.d.)  over the REC$[\gamma,\delta]$; Bob receives $\bY$. \\
\noindent (C2). Bob chooses a hash function $G_1\sim \text{Unif}\left(\mathcal{G}_1\right)$, and sends the description of $G_1$ to Alice over the noiseless link.\\ 
\noindent (C3). Alice computes $G_1(\bX)$ and sends it to Bob over the noiseless link.\\
\noindent (C4). Bob picks another hash function $G_2\sim\text{Unif}\left(\mathcal{G}_2\right)$, and sends its description to Alice over the noiseless link.\\ 
\noindent (C5). Alice computes the hash $G_2(\bX)$ and sends it over the noiseless link to Bob.\\
\noindent (C6). Alice chooses an extractor function $\text{Ext}\sim\text{Unif}\left(\mathcal{E}\right)$ and sends\footnote{In the following expression, operator $\oplus$ denotes component-wise XOR.} $Q = c \oplus \text{Ext}(\bX)$ and the description of $\text{Ext}$ to Bob over the noiseless link.\\
\indent $\bullet$\emph{Reveal phase:} Alice proceeds as follows: \\
\noindent \ajb{(R1). Having received $\bY=\by$, Bob creates list $\cL(\by)$ of vectors given by:\footnote{Here the parameter $\alpha_1>0$ is chosen appropriately small.}
\begin{IEEEeqnarray*}{rCl}
\cL(\by):=\{\bx\in \{0,1\}^n: n(\delta -\alpha_1) \leq d_H(\bx,\by) \leq n(\delta +\alpha_1) \}.
\end{IEEEeqnarray*}
}
\noindent (R2). Alice announces  $(\td{c},\td{\bx})$ to Bob over the noiseless link.\\
\noindent (R3). Bob accepts $\td{c}$ if all the following four conditions are satisfied: $(i)$ $\td{\bx}\in\cL(\by)$, $(ii)$ $g_1(\tilde{\bx})=g_1({\bx})$, $(iii)$ $g_2(\tilde{\bx})=g_2({\bx})$ and $(iv)$ $\tilde{c}=q~\oplus \text{ext}(\tilde{\bx})$. Else, he rejects $\td{c}$ and outputs `0'.

We now analyse and prove the security guarantees in detail for the above defined $(n,R)$-commitment scheme:
%

[1] \underline{\emph{$\epsilon-$sound:}} For our protocol to be $\epsilon$-sound, it is sufficient to show that  $\bbP\left(\bX\not\in \cL(\bY)\right)\leq \epsilon$ when both the parties, Alice and Bob, are honest; the proof of this fact follows from classic Chernoff bounds. We skip the details.

[2] \underline{\emph{$\epsilon$-concealing:}}
It is known that a positive rate commitment protocol is $\epsilon-$concealing, where $\epsilon>0$ is \emph{exponentially decreasing} in blocklength $n$, if it satisfies the \emph{capacity-based secrecy} (cf.~\cite[Def.~3.2]{damgard1998statistical}) and vice versa. We use a well established relation between \emph{capacity-based secrecy} and the \emph{bias-based secrecy} (cf.~\cite[Th.~4.1]{damgard1998statistical}) to prove that our protocol is $\epsilon$-concealing. 

  To begin,  we  prove that our protocol satisfies bias-based secrecy by essentially proving the perfect secrecy of the key $\text{Ext}(\bX)$; here we crucially use the  \emph{leftover hash} lemma. Several versions of this lemma exists (cf.~\cite{impagliazzo1989pseudo,glh,haastad1999pseudorandom} for instance); we use the following:
\begin{lemma}\label{lem:glh}
Let $\mathcal{G}=\{G:\{0,1\}^n\rightarrow \{0,1\}^l\}$ be a family of universal hash functions. Then, for any hash function $G$ chosen  uniformly at random from $\mathcal{G}$, and $W$
\begin{align*}
    \|(P_{G(W),G}-P_{U_l,G})\| \notag&\leq \frac{1}{2}\sqrt{2^{ -H_{\infty}(W)} 2{^l}}
\end{align*}
where $U_l\sim\text{Unif}\left(\{0,1\}^l\right).$
\end{lemma}
%
%
We then establish the following lower bound:
\begin{lemma}\label{lem:smooth:min:entropy}
For any $\epsilon_1>0, \zeta>0$ and $n$ sufficiently large, 
\begin{align}
	H_{\infty}^{\epsilon_1}&(\bX|\bY,G_1(\bX),G_1,G_2(\bX),G_2)\notag\\
	&\stackrel{}{\geq} { n(H(\delta)-\zeta-H(\kappa)-\beta_1-\beta_2)}-\log(\epsilon_1^{-1})\label{eq:h:inf:1}
    \end{align}
\end{lemma}
The proof appears in Appendix~\ref{app:smooth:min:entropy}. Next, we use Lemma~\ref{lem:glh} to show that the distribution of the secret key $\text{Ext}(\bX)$ is statistically close to  a uniform distribution thereby achieving bias-based secrecy. Let us fix $\epsilon_1:=2^{-n\alpha_2}$, where $\alpha_2>0$ is an arbitrary small constant.
We make the following correspondence in  Lemma~\ref{lem:glh}: $G\leftrightarrow \text{Ext}$, $W\leftrightarrow \bX$ and  $l\leftrightarrow nR$
to get the following:
\begin{align}
    \|&P_{\text{Ext}(\bX),\text{Ext}}-P_{U_l,\text{Ext}}\| \notag \\
    &\stackrel{(a)}{\leq} \frac{1}{2}\sqrt{2^{ -H_{\infty}(\bX)} 2{^{nR}}}\notag\\
		&\stackrel{(b)}{\leq} \frac{1}{2}\sqrt{2^{ -H_{\infty}(\bX|\bY,G_1(\bX),G_1,G_2(\bX),G_2)} 2{^{nR}}}\notag\\
    &\stackrel{(c)}{\leq} \frac{1}{2}\sqrt{2^{-n(H(\delta)-\zeta-H(\kappa)-\beta_1-\beta_2-\alpha_2)} 2^{n(H(\delta)-H(\kappa)-\beta_3))}}\notag\\
    &= \frac{1}{2}\sqrt{2^{n(\zeta+\beta_1+\beta_2+\alpha_2-\beta_3))}}\notag\\
    &\stackrel{(d)}{\leq} 2^{-n\alpha_3}  \label{eq:sd:1}
\end{align}
where, $\alpha_3 > 0$ and $n$ is sufficiently large.
Here,\begin{enumerate}[(a)]
\item follows directly from the leftover hash lemma (cf.~Lemma~\ref{lem:glh})
\item follows from the fact that conditional min-entropy bounds min-entropy.
\item follows from (\ref{eq:h:inf:1}) and noting that the choice of 2-universal hash function $\text{Ext}$ is random and  uniform  from the set  ${\mathcal{E}}:\{0,1\}^n \rightarrow \{0,1\}^{n(H(\delta)-H(\kappa)-\beta_3)}$.
\item follows from noting that $\beta_3$ is chosen such that $\zeta+\beta_1+\beta_2+\alpha_2-\beta_3<0$; here, we note that $\alpha_2$ is an arbitrarily chosen (small enough) constant, and $\zeta>0$  can be made arbitrarily small for $n$ sufficiently large. As such, a choice of $\beta_3>\beta_1 +\beta_2$ is sufficient.
\end{enumerate}
From~\eqref{eq:sd:1} and Lemma~\ref{lem:glh}, it follows that we can extract $n(H(\delta)-H(\kappa)-\beta_3)$ almost uniformly random bits which proves the security of the secret key; this guarantees that our commitment protocol satisfies bias-based secrecy (cf.~\cite[Def.~3.1]{damgard1998statistical}).  Recall from our discussion earlier (see also~\cite[Th.~4.1]{damgard1998statistical}) that bias-based secrecy under \emph{exponentially decaying} statistical distance, as in~\eqref{eq:sd:1}, implies capacity-based secrecy; hence, it follows that for $n$ sufficiently large,  $I(C;\view_B)\leq \epsilon$ and our protocol is $\epsilon$-concealing.

[3] \underline{\emph{$\epsilon$-binding:}}
%
%
\removed{
\begin{figure}[!ht]
        \centering
        \includegraphics[scale=0.5]{bind2.jpg}
        \caption{Alice's cheating strategy:  set REC$[\gamma,\delta]$ to BSC($s$), $s\in[\gamma,\delta]$ and send $\bx$ in the commit phase, and then reveal  $\bx'\neq \bx$ in the reveal phase. It can be shown that the `best' such choice for Alice is to fix $s=\gamma$.}
        \label{fig:my_label}
\end{figure}
}
 Let us assume that a dishonest Alice sets the crossover probability of the REC$[\gamma,\delta]$ to $s\in[\gamma,\delta]$; let us define $\kappa_s:=\frac{\delta-s}{1-2s}$. Note that $\kappa=\kappa_{\gamma}=\frac{\delta-\gamma}{1-2\gamma}$.
 Let $\bX=\bx$ be the transmitted bit string and $\bY=\by$ be the  bit string received by Bob's over the BSC($s$). Alice can cheat successfully by confusing Bob in the reveal phase only if she can find two distinct bit strings $\bx'$ and $\td{\bx}$ such that (i) $\bx',\td{\bx} \in \cL(\by)$, and (ii) $\bx'$, $\td{\bx}$ pass the two rounds of sequential random hash exchange challenge (w.r.t hash functions $G_1(\cdot)$ and $G_2(\cdot)$). 
Let $\cA$ denote all such candidate vectors that appear in Bob's list (\emph{prior} to the hash challenges) that Alice can use to confuse Bob; the following claim shows that $\cA$ can be exponentially large.
\begin{claim}\label{claim:A:set}
Given any $\eta>0$, for $n$ sufficiently large
\begin{equation}
|\cA|\leq 2^{n(H(\kappa)+\eta)}\label{eq:A}
\end{equation}
\end{claim}
The proof appears in Appendix~\ref{app:A:set}. Note that, essentially, we can conclude that the choice of $s=\gamma$ is the `best' choice for a cheating Alice (such a choice maximizes $|\cA|$), i.e., Alice can be no worse than when it fixes the REC to a BSC($\gamma$). We will choose  $0<\eta<\beta_1$ later (cf. Claim~\ref{claim:step:1}). 

We now show that our choice of hash functions $G_1(\cdot)$ and $G_2(\cdot)$ allows us to essentially `trim' down this set $\cA$ of `confusable' vectors all the way down to none.  Recall that Alice's choice in the commit phase is $\bx$. For a given hash value $h_1\in\{0,1\}^{n(H(\kappa) + \beta_1)}$ sent by Alice, let 
\begin{equation}
I_i(h_1):=\begin{cases}
1& \mbox{ if } G_1(\bx_i)=G_1(\bx)=h_1 \\
0& \mbox{ otherwise.}
\end{cases}
\end{equation}
Also, let 
\begin{align}
I(h_1):=\sum_{i=1}^{|\cA|} I_i(h_1)
\end{align}
 denotes the total number of hash collisions with hash value $h_1$.
Then, the following holds when $0<\eta<\beta_1$:
\begin{claim}\label{claim:step:1}
$\mathbb{P}\left(\exists h_1\in \{0,1\}^{n(H(\kappa)+ \beta_1)}: I(h_1)>8n+1\right)\rightarrow 0$ exponentially in $n$ as $n\rightarrow\infty.$
\end{claim}
This implies that the size of the `confusable' set \emph{after} the first hash challenge via $G_1$ for any $h_1$ is larger that $8n+1$ with exponentially small probability (in block length $n$). 

Conditioned on the event $I(h_1)<8n+1$, $\forall h_1$, which occurs with high probability (w.h.p.), we now analyse the size of the `confusable' set \emph{after} the second hash challenge via $G_2$; let $\mathcal{F}_{h_1}$ denote this set of `confusable' vectors after the second hash challenge for a given $h_1$. We prove the following claim (proof in Appendix~\ref{app:step:2}):
\begin{claim}\label{claim:step:2}
For every $h_1\in\{0,1\}^{n(H(\kappa) + \beta_1)}$, we have for $n$ sufficiently large
\begin{align}
\mathbb{P}&\left(\exists \bx\neq \bx'\in\mathcal{F}_{h_1}:G_2(\bx)=G_2(\bx')\big| I(h_1)\leq 8n+1\right)\notag\\
&\leq 2^{-n\frac{\beta_2}{2}} \label{eq:bind:1}
\end{align}
\end{claim}
As~\eqref{eq:bind:1} holds for every $h_1$, and noting that\footnote{Recall that $\beta_2>0$ is a fixed parameter in our protocol.} $\beta_2>0$, we now choose $n$ large enough to prove that our commitment protocol is $\epsilon-$binding.

%% file: conclusion.tex
In summary, we characterized the commitment capacity of the RECs; this settles affirmatively, a recent conjecture (cf.~\cite{crepeau2020commitment}) on the same. A key contribution in this work is our general converse which analyses a specific cheating strategy of a dishonest Alice to establish a tight rate upper bound.

Coupled with existing results for UNCs and ECs (cf.~\cite{crepeau2020commitment,khurana2016secure}), our result shows that for a fixed set of parameters $0<\gamma<\delta<1/2$, the commitment capacities can be ordered as follows: $\mathbb{C}_{EC}>\mathbb{C}_{REC}>\mathbb{C}_{UNC}$.\footnote{For the case when $\gamma=\delta$, all channels default to a BSC($\delta$) which offers the highest throughput $H(\delta).$} 
This ordering implies that the commitment throughput degradation  in RECs and ECs (vis-\`{a}-vis a fixed BSC($\delta$)) is `not symmetric' in the \emph{one-sided elasticity} available exclusively to the committer   and the receiver  resp. in those models. In particular, a dishonest committer  is more limiting (w.r.t. commitment throughput) than a dishonest receiver. 

Both RECs and ECs are unreliable channels with one-sided elasticity, where exactly one amongst the committer and the receiver can alter the channel non-trivially when dishonest, but not both. Crucially, when both parties are honest the channel defaults to a classic BSC. Then, we ask the following question: can one define unreliable elastic channels with \emph{two-sided elasticity}, where both the committer and receiver, when dishonest, can  alter the channel non-trivially whilst keeping the honest party unaware of the same? 
Furthermore, is it possible to model  \emph{unequal} committer-side and receiver-side elasticities, say $\gamma_A<\delta$ and $\gamma_B<\delta$ respectively? We answer the above questions affirmatively and propose the study of such two-sided elastic channels via the general framework of a \emph{general elastic channel} GEC$[\gamma_A,\gamma_B,\delta]$, where $0<\gamma_A,\gamma_B<\delta<1/2$. For a `symmetric' instance of such a general elastic channel, i.e., for a GEC$[\gamma,\gamma,\delta],$ we conjecture that its commitment capacity equals that of the REC$[\gamma,\delta].$ This conjecture stems from our understanding of the committer-receiver asymmetry (vis-\`{a}-vis commitment capacity) over channels with  elasticity, albeit under one-sided elasticity, which we presented in this work. We believe that for the GEC$[\gamma,\gamma,\delta]$ commitment capacity characterization, the REC$[\gamma,\delta]$-like converse is tight though the main  bottleneck is the achievability protocol. We leave the capacity of the GEC$[\gamma,\gamma,\delta]$ (and that of the more general GECs) as an open problem.

Finally, the general focus of this work was channels with elasticity, which as the name suggests, model `unreliability' in channels via \emph{elasticity.} Another pertinent, though different form of unreliability in channels appears via `compound' channels. In a compound-setting,  the channel observed by two \emph{honest}  parties is fixed but not known to them (unlike in standard elastic channels where it is fixed and known); instead a set comprising potential candidate channels (including the instantiated one) is known to parties \emph{a priori.} The UNC is a classic example of a channel which combines both these forms of unrealiability, viz., elasticity and compound-nature, albeit in a symmetric manner. 
Seen from this perspective, the significantly lower commitment capacity over UNCs (w.r.t. RECs and ECs) can be attributed to the compound-nature of the UNC (when both parties are honest) in addition to the two-sided elasticity (when parties are dishonest); see Fig.~\ref{fig:capacity:curves}. In a future work, we seek to explore general unreliable channels combining both, the  elasticity and compound-nature, in channels. We believe that such a framework will significantly generalize the scope of study over unreliable channels.

%% file: appendix.tex
\subsection{Proof of Lemma~\ref{lem:fano:bindsound}}\label{app:fano:lemma}
We use the fact that $\mathscr{P}_n$ is $\epsilon_n$-sound and $\epsilon_n$-binding in this proof; furthermore, 
we can show that every protocol $\mathscr{P}_n$ can essentially recover the commit string under a `noisy' version $\bZ$  of $\bX$ coupled with $\bY$ (we use Fano's inequality here). 

Let us define\footnote{Although Bob's test $T$ is a randomized test, it can be shown that one can construct from $T$ a deterministic test with essentially the same soundness and bindingness performance. Hence, for the rest of the converse, we consider that Bob's test is a deterministic function; as such, $\hat{c}$ is well defined for such a deterministic test.} 
%
$\hat{c}(\bZ,V_B):=\arg \max_{c\in[2^{nR}]} T(c,\bZ,V_B).$
%
Here we crucially use the fact that for Alice's assumed cheating strategy where she fixes the channel to Bob as  BSC($s$), $s\in[\gamma,\delta]$, the effective channel from $Z$ to $Y$ is a BSC with crossover probability $\kappa_s\otimes s=\delta$ under every $s\in[\gamma,\delta]$. 

We now bound $\mathbb{P}(\hat{C}\neq C)$, where $\hat{C}=\hat{c}(\bZ,V_B)$.
As the code is $\epsilon_n$-binding, it follows that
\begin{align}
\mathbb{P}\left(T(\bar{c},\bar{\bX},V_B)=1 \quad\& \quad T(\hat{c},\hat{\bX},V_B)=1 \right)\leq \epsilon_n\notag
\end{align}
for any two distinct $(\bar{c},\bar{\bX})$ and $(\hat{c},\hat{\bX})$ such that $\bar{c}\neq \hat{c}.$
Furthermore, as the code is $\epsilon_n$-sound, 
\begin{align}
\mathbb{P}\left(T(c,\bZ,V_B)=1 \right)\geq 1-\epsilon_n. \notag
\end{align}
where we crucially use the fact that $Z$ to $Y$ is a BSC($\delta$) channel. Note that for the converse, we assume an \emph{averaged} (over commit strings $C$) soundness criterion, where we replace the `$\max$' in~\eqref{eq:sound} with an average over $C$.\footnote{This is a stronger converse as impossibility under the \emph{average} criterion implies impossibility over the \emph{maximal} criterion in~\eqref{eq:sound}.}
Thus, for the given decoder, we then have
\begin{align*}
\mathbb{P}(\hat{C}\neq C)&=\mathbb{P}(\hat{C}=0)+\mathbb{P}(\hat{C}\neq C|\hat{C}\neq 0)\\
												  &\leq \epsilon_n+ \epsilon_n\\
													&\leq 2\epsilon_n.
\end{align*}
where in the penultimate inequality, the first part follows from noting that  $\mathscr{P}_n$ is $\epsilon_n$-binding, and the second part follows from the fact that conditioned on $\mathscr{P}_n$ being $\epsilon_n$-binding, the probability that $\hat{C}=\hat{C}(\bZ,V_B)$ is different from $C$ is at most $\epsilon_n$ due to $\mathscr{P}_n$ being $\epsilon_n$-sound.

We now use Fano's inequality (cf.~\cite{elgamal-kim}) to bound the conditional entropy.
\begin{align*}
H(C|\bZ,V_B)&\leq 1+ \mathbb{P}(\hat{C}\neq C) nR\\
					 &\leq n\epsilon'_n  \quad 
\end{align*}
where $\epsilon'_n(\epsilon):=\frac{1}{n}+2\epsilon_n R\rightarrow 0$ as $\epsilon_n\rightarrow 0.$ 
\subsection{Proof of Lemma~\ref{lem:smooth:min:entropy}}\label{app:smooth:min:entropy}
Before we start with the proof, we recap (without proof) a few well known results.
\begin{claim}[Min-entropy~\cite{chain1}]\label{claim:min:entropy}
 For any $0 \leq \mu,\mu',\mu_1,\mu_2 <1$ and any set of jointly distributed random variables $(X, Y, W)$, we have 
\begin{IEEEeqnarray}{rCl}
&&H_{\infty}^{\mu+\mu^{'}}(X,Y|W)-H_{\infty}^{\mu^{'}}(Y|W)\notag\\ 
&&\geq H_{\infty}^{\mu}(X|Y,W)\label{eq:min:1}\\ 
&&\geq H_{\infty}^{\mu_1}(X,Y|W)-H_0^{\mu_2}(Y|W)-\log\left[\frac{1}{\mu-\mu_1-\mu_2}\right]\label{eq:min:2}
\end{IEEEeqnarray}
\end{claim}
\begin{claim}[Max-entropy~\cite{chain1, chain2}]\label{claim:max:entropy}
For any $0 \leq \mu,\mu',\mu_1,\mu_2 <1$ and any set of jointly distributed random variables $(X, Y, W)$, we have 
\begin{IEEEeqnarray}{rCl}
&&H_{0}^{\mu+\mu^{'}}(X,Y|W)-H_{0}^{\mu^{'}}(Y|W) \notag\\
&&\leq H_{0}^{\mu}(X|Y,W)\label{eq:max:1}\\
&&\leq H_{0}^{\mu_1}(X,Y|W)-H_{\infty}^{\mu_2}(Y|W)+\log\left[\frac{1}{\mu-\mu_1-\mu_2}\right]\label{eq:max:2}
\end{IEEEeqnarray}
\end{claim}
Now consider the following for any $\epsilon_1>0$:
\begin{align}
	&H_{\infty}^{\epsilon_1}(\bX|\bY,G_1(\bX),G_1,G_2(\bX),G_2)\notag\\
	&\stackrel{(a)}{\geq} H_{\infty}(\bX,G_1(\bX),G_2(\bX)|\bY,G_1,G_2)\notag\\
	&\hspace{10mm}-H_{0}(G_1(\bX),G_2(\bX)|\bY,G_1,G_2)-\log(\epsilon_1^{-1})\notag\\
	&\stackrel{(b)}{=} H_{\infty}(\bX|\bY,G_1,G_2)+H_{\infty}(G_1(\bX),G_2(\bX)|\bY,G_1,G_2,\bX)\notag\\
	&\hspace{10mm}-H_{0}(G_1(\bX),G_2(\bX)|\bY,G_1,G_2)-\log(\epsilon_1^{-1})\notag\\
	&\stackrel{(c)}{=} H_{\infty}(\bX|\bY,G_1,G_2)\notag\\
	&\hspace{10mm}-H_{0}(G_1(\bX),G_2(\bX)|\bY,G_1,G_2)- \log(\epsilon_1^{-1})\notag\\
	&\stackrel{(d)}{=} H_{\infty}(\bX|\bY)-H_{0}(G_1(X),G_2(X)|\bY,G_1,G_2)- \log(\epsilon_1^{-1})\notag\\
	&\stackrel{(e)}{\geq}  H_{\infty}(\bX|\bY)-H_{0}(G_1(\bX)|G_2(\bX),\bY,G_1,G_2)\notag\\
	&\hspace{10mm}-H_{0}(G_2(\bX)|\bY,G_1,G_2)-\log(\epsilon_1^{-1})\notag\\
	&\stackrel{(f)}{\geq}  (H(\bX|\bY)-\zeta')-H_{0}(G_1(\bX)|G_2(\bX),\bY,G_1,G_2)\notag\\
	&\hspace{10mm}-H_{0}(G_2(\bX)|\bY,G_1,G_2)-\log(\epsilon_1^{-1})\notag\\
	&\stackrel{(g)}{\geq} { n(H(\delta)-\zeta)}-n(H(\kappa) + \beta_1+\beta_2)-\log(\epsilon_1^{-1})\notag\\
	&\stackrel{}{=} { n(H(\delta)-\zeta-H(\kappa)-\beta_1-\beta_2)}-\log(\epsilon_1^{-1})
    \end{align}
where we have
\begin{enumerate}[(a)]
\item from the chain rule for smooth min-entropy; see Claim~\ref{claim:min:entropy} and substitute $\mu=\epsilon_1$, $\mu_1=0$ and $\mu_2=0$ in~\eqref{eq:min:2}.
\item from the chain rule for min-entropy; see Claim~\ref{claim:min:entropy} and substitute $\mu=0$ and $\mu'=0$ in~\eqref{eq:min:1}.
\item from the fact that $G_1(\bX)$ and $G_2(\bX)$  are deterministic functions of $G_1$, $G_2$ and $\bX$. 
\item by the Markov chain  $\bX \leftrightarrow \bY \leftrightarrow (G_1,G_2)$.
\item from the chain rule for max-entropy; see Claim~\ref{claim:max:entropy} and  substitute $\mu=0$ and $\mu'=0$ in~\eqref{eq:max:1}. 
\item from~\cite[Th.~1]{nascimento-barros-t-it2008} which allows us to lower bound $H_{\infty}(\bX|\bY)$ in terms of $H(\bX|\bY)$  (via an appropriate smooth-min-entropy quantity); here $\zeta>0$  can be made
arbitrarily small for $n$ sufficiently large
\item by noting that the crossover probability is $\delta$ and from definition of max-entropy (also noting that the  range of $G_1$ and $G_2$ is $\{0,1\}^{n(H(\kappa) + \beta_1)}$ and $\{0,1\}^{n\beta_2}$ respectively).
\end{enumerate}
\subsection{Proof of Claim~\ref{claim:A:set}}\label{app:A:set}
From the definition of $\cA$, we have
\begin{align}
|\cA|&\stackrel{(a)}{\leq} 2^{n(H(\kappa_s)+\eta)}\notag\\
 &\stackrel{(b)}{\leq} 2^{n(H(\kappa)+\eta)}
\end{align}
where 
\begin{enumerate}[(a)]
\item follows from noting that an honest Bob will accept a vector $\bx'$ if $d_H(\bx'\by)\in[n(\delta-\alpha_1),n(\delta+\alpha_1)]$; since Alice has fixed the REC$[\gamma,\delta]$ to a BSC($s$), the total number of such vectors are at most $2^{n(H(\kappa_s)+\eta)}$, where $\eta>0$ choice can be arbitrary, for $n$ sufficiently large. 
\item follows from noting that $\kappa_s\leq\kappa=\frac{\delta-\gamma}{1-2\gamma}<1/2.$ 
\end{enumerate}
This concludes the proof of the claim.
\subsection{Proof of Claim~\ref{claim:step:1}}\label{app:step:1}
Recall that $G_1\sim \text{Unif}\left(\mathcal{G}_1\right)$. Then,
\begin{align}
\mathbf{E}_{G_1}[I(h_1)]
&\stackrel{(a)}{\leq}\sum_{i=1}^{|\cA|}2^{-(n(H(\kappa)+\beta_1))} \notag \\
&\stackrel{(b)}{\leq}2^{n(\eta-\beta_1)}\notag\\
&\stackrel{(c)}{\leq}2^{-n\td{\beta}_1}\label{eq:I}
\end{align} 
which is independent of $h_1$. Here $(a)$ follows from the definition of $\mathcal{G}_1$, $(b)$ follows from~Claim~\ref{claim:A:set} and noting that $\beta_1> \eta$; letting $\td{\beta}_1:=\beta_1-\eta>0$  gives us $(c)$. Note that for $n$ sufficiently large, we have $\mathbb{E}[I(h_1)]\leq 1$, $\forall h_1$.

We now need the following result by Rompel~\cite{rompel} to proceed:
\begin{lemma}[~\cite{rompel}]\label{lem:rompel}
	Let $X_1,X_2,X_3....X_m\in [0,1]$ be $k$-wise independent random variables, where $k$ is an even and positive integer. Let  $X:=\sum_{i=1}^{m} X_i$, $\mu:=\mathbf{E}[X]$, and $\Delta>0$ be a constant. Then,
	\begin{align}
	\mathbb{P}\left(|X-\mu|>\Delta\right)<O\left(\left(\frac{k\mu+k^2}{\Delta^2}\right)^{k/2}\right)    
	\end{align}
\end{lemma}
We now make the following correspondence: $k\leftrightarrow 4n$, $\Delta\leftrightarrow 2k = 8n$. Then, using the union bound, we get:
\begin{align}
\mathbb{P}&\left(\exists h_1\in \{0,1\}^{n(H(\kappa)+ \beta_1)}: I(h_1)>8n+1\right) \\
&\leq  \sum_{h_1\in \{0,1\}^{n(H(\kappa)+ \beta_1)} } \mathbb{P}\left(I(h_1)>8n+1\right)\\
&\stackrel{(a)}{\leq}  2^{n(H(\kappa)+\beta_1)}O\Big(\Big(\frac{k\mu+k^2}{\Delta^2}\Big)^{k/2}\Big) \notag\\
&\stackrel{(b)}{\leq} 2^{n(H(\kappa)+\beta_1)} O\Big(\Big(\frac{1+k}{4k}\Big)^{k/2}\Big) \notag\\
&<2^{n(H(\kappa)+\beta_1)} O(2^{-k/2}) \notag\\
&=2^{n(H(\kappa)+\beta_1)} O(2^{-n}) \label{eq:aa} 
\end{align}
where we have
\begin{enumerate}[(a)]
\item from Lemma~\ref{lem:rompel}
\item by noting that for $n$ sufficiently large,  $\mu=\mathbb{E}[I(h_1)]\leq 1$, $\forall h_1$, and making the correspondence $\Delta\leftrightarrow 2k$.
\end{enumerate}
Now note that~\eqref{eq:aa} tends to zero exponentially fast as we have $(H(\kappa)+\beta_1)<1$.  This completes the proof of claim.
\subsection{Proof of Claim~\ref{claim:step:2}}\label{app:step:2}
Recall the definition of $\mathcal{F}_{h_1}$, and let $\mathcal{F}:=\max_{h_1} \mathcal{F}_{h_1}$. Note that $|\mathcal{F}|\leq 8n+1$. Noting that $G_2\sim\text{Unif}\left(\mathcal{G}_2\right)$, where $\mathcal{G}_2=\{g_2:\{0,1\}^n \rightarrow \{0,1\}^{n\beta_2}\}$, we have for every $h_1\in\{0,1\}^{n(H(\kappa) + \beta_1)}$,
\begin{align}
\mathbb{P}&\left(\exists \bx\neq \bx'\in\mathcal{F}_{h_1}:G_2(\bx)=G_2(\bx')\big| I(h_1)\leq 8n+1\right)\notag\\
&\stackrel{(a)}{\leq} {\mathcal{F}\choose {2}} \mathbb{P}\left(G_2(\bx)=G_2(\bx')\right) \notag\\
&\stackrel{(b)}{\leq} {{8n+1}\choose {2}}2^{-n\beta_2} \notag\\
&< (8n+1)(8n)2^{-n\beta_2} \notag\\
&\leq 2^{-n\frac{\beta_2}{2}} \hspace{5mm} \text{ for $n$ large enough}
\end{align}
where $(a)$ follows from the definition of $\mathcal{F}$, and using the union bound (on distinct pairs of vectors in $\mathcal{F}$); we get $(b)$ from the definition of $\mathcal{G}_2$. This completes the proof of the claim.